%% file: higgs+jet.tex
\documentclass[prd,superscriptaddress,twocolumn, 
preprintnumbers,showpacs]{revtex4-1} 
\pdfoutput=1
\usepackage{amssymb,amsmath,graphicx,epsfig} 
\usepackage{color}
\usepackage{float}
\usepackage{xcolor}
\definecolor{darkgreen}{rgb}{0,0.4,0}
\definecolor{darkred}{rgb}{0.7,0,0}
\definecolor{darkblue}{rgb}{0,0,0.5}
\usepackage[pdftex,
                colorlinks=true,
                urlcolor=darkblue,
                citecolor=darkgreen,
                linkcolor=darkred]{hyperref}
\usepackage{base}
\def\beq{\begin{equation}} 
\def\eeq{\end{equation}} 
\def\bea{\begin{eqnarray}} 
\def\eea{\end{eqnarray}} 
\def\nn{\nonumber}

\newcommand{\Lcal}{\mathcal{L}}

\begin{document} 
\title{The Dimension Six Triple Gluon Operator in Higgs+Jet Observables}
\author{Diptimoy Ghosh} 
\email{diptimoy.ghosh@roma1.infn.it} 
\affiliation{INFN, Sezione di Roma, Piazzale A. Moro 2, I-00185 Roma, Italy} 
\author{Martin Wiebusch} 
\email{martin.wiebusch@durham.ac.uk} 
\affiliation{Institute for Particle Physics Phenomenology, 
Department of Physics, Durham University, Durham CH1 3LE, United Kingdom} 
\begin{abstract} 
  Recently a lot of progress has been made towards a full classification of new
  physics effects in Higgs observables by means of effective dimension six
  operators. Specifically, Higgs production in association with a high
  transverse momentum jet has been suggested as a way to discriminate between
  operators that modify the Higgs-top coupling ($O_t$) and operators that induce
  an effective Higgs-gluon coupling ($O_g$)---a distinction that is hard to
  achieve with signal strength measurements alone.  With this article we would
  like to draw attention to another source of new physics in Higgs+jet
  observables: the triple gluon operator $O_{3g}$ (consisting of three factors
  of the gluon field strength tensor). We compute the distortions of kinematic
  distributions in Higgs+jet production at a \unit{14}{TeV} LHC due to $O_{3g}$
  and compare them with the distortions due to $O_t$ and $O_g$. We find that the
  transverse momentum distributions alone can not discriminate between $O_{3g}$
  and $O_g$ if the coefficient of the operator $O_t$ treated as an unknown
  parameter. We further show that the jet rapidity and the difference between
  the Higgs and jet rapidity are well suited to remove this new
  degeneracy. Using rough estimates for the expected bounds we find that allowed
  distortions in kinematic distributions due to $O_g$ are of similar size as
  those due to $O_{3g}$. We conclude that a full analysis of new physics in
  Higgs+jet observables must take the contributions from $O_{3g}$ into account.
\end{abstract}
\keywords{Higgs boson, effective operators, quantum chromodynamics}
\pacs{12.60.Fr,12.38.Bx,14.80.Bn}
\preprint{IPPP/14/96\\DCTP/14/192}
\maketitle
\section{Introduction}
\label{sec:intro}

The discovery of a Higgs resonance whose properties match those of the Standard
Model (SM) Higgs boson and the absence of any clear signs of physics beyond the
SM in other collider searches suggests that, contrary to many people's
expectations, the SM remains valid at energy scales significantly larger than
the electroweak scale. In such a situation the language of effective field
theory allows us to systematically classify all possible effects that new
physics can have on low energy observables in terms of higher dimensional
operators. Recently a lot of effort has been dedicated to defining a complete
basis of dimension six operators \cite{Grzadkowski:2010es, Contino:2013kra,
  Jenkins:2013zja, Jenkins:2013wua} and analysing experimental constraints on
these operators \cite{Corbett:2012ja, Dumont:2013wma, Elias-Miro:2013mua,
  Pomarol:2013zra, Almeida:2013jfa, Alonso:2013hga, Ellis:2014dva,
  deBlas:2014ula, Ellis:2014jta}.

Determining the Wilson coefficients of all 59 independent dimension six
operators clearly requires a large number of experimental
observables. Specifically, using only Higgs signal strength measurements as
provided by the ATLAS and CMS collaborations \cite{ATLAS-CONF-2014-009,
  CMS-PAS-HIG-14-009}, it is very difficult to discriminate between the operator
$O_t$ that modifies the the $Ht\bar t$ coupling and the operator $O_g$ that
generates an effective $Hgg$ coupling. (The $Ht\bar t$ coupling is only probed
directly in Higgs production in association with a $t\bar t$ pair, and this
process is difficult to measure due to its small rate and complicated final
state.)  As discussed in \cite{Harlander:2013oja, Banfi:2013yoa, Azatov:2013xha,
  Grojean:2013nya, Schlaffer:2014osa} the degeneracy can be broken by measuring
kinematic distributions in Higgs production in association with a high $p_T$
jet. Further studies of kinematic distributions in Higgs+jet production in the
presence of dimension six operators were performed in \cite{Gainer:2014hha,
  Buschmann:2014twa, Buschmann:2014sia} and higher order corrections were
considered in \cite{Englert:2014cva, Dawson:2014ora}.

All the above-mentioned studies concentrate on dimension six operators that
contain the SM Higgs doublet. With this article we would like to raise awareness
for the fact that Higgs production in association with a jet also receives
contributions from the triple gluon operator $O_{3g}$ (consisting of three
factors of the gluon field strength tensor). Following the classification of
\cite{Elias-Miro:2013mua} this operator is ``loop suppressed'' compared to the
operators involving the SM Higgs doublet. However, a truely model independent
analysis of new physics effects in Higgs+jet production should take these
contributions into account. At the very least the Wilson coefficient $C_{3g}$ of
$O_{3g}$ should appear as a nuisance parameter in the extraction of the Wilson
coefficients of other operators from Higgs+jet observables. Constraints from top
pair production on $C_{3g}$ were discussed in \cite{Cho:1994yu}. Experimental
results for boosted top-pair production from the 7 and \unit{8}{TeV} LHC runs
are presented in \cite{ATLAS-CONF-2014-057} but not interpreted in the context
of dimension six operators. Here we study the modifications of kinematic
distributions in Higgs+jet production due to the triple gluon operator and
compare them with those induced by operators involving the Higgs doublet. We
find that $p_T$ distributions alone are insufficient to disentangle the
contributions from $O_t$, $O_g$ and $O_{3g}$, and that this degeneracy can be
broken by considering rapidity distributions.  In particular the rapidity
distribution of the extra jet---although incapable of discriminating between
modified $Ht\bar t$ and effective $Hgg$ couplings---turns out to be sensitive to
contributions from $O_{3g}$.

The presented results are intended as a motivation for more detailed
analyses. Hence all our calculations are done at leading order and we do not
include parton shower or detector effects. We also postpone the estimation of
the experimental sensitivity to the Wilson coefficients to a future
publication. Note that, since we are only interested in the shapes of kinematic
distributions the inclusion of universal QCD $K$-factors is unnecessary for this
paper and a fully differential next-to-leading order calculation is clearly
beyond its scope.

\section{Effective Operators in
  \texorpdfstring{$\boldsymbol{gg\to Hg}$}{gg->Hg}}
\label{sec:op}

We consider the effective dimension six Lagrangian
\begin{eqnarray}\label{eq:L6}
  \Lcal_6 =
    \frac{C_g}{\Lambda^2}O_g
  + \frac{C_{3g}}{\Lambda^2}O_{3g}
  &+& \left(\frac{C_{t}}{\Lambda^2}O_{t} + \rm h.c.\right) \nn \\
  &+& \left(\frac{C_{b}}{\Lambda^2}O_{b} + \rm h.c.\right) \, ,  
\end{eqnarray}
where $\Lambda$ is the scale of new physics, $C_g$, $C_{3g}$,
$C_{t}$, and $C_{b}$ are dimensionless Wilson coefficients and
\begin{align}
  O_g &= \Phi^{\dagger}\Phi\,G^a_{\mu\nu} G^{\mu\nu a}
  \,\,, &
  O_t &= Y_t(\Phi^\dagger\Phi)(\bar Q_{3L} t_R \tilde\Phi)
  \,\, ,\nonumber\\
  O_{3g} &= f^{abc} ?G^{a\mu}_\nu? ?G^{b\nu}_\rho? ?G^{c\rho}_\mu?
  \,\, ,&
  O_b &= Y_b(\Phi^\dagger\Phi)(\bar Q_{3L} b_R \Phi)
  \,\, .\label{eq:op}
\end{align}
Here $\Phi$ denotes the SM Higgs doublet, $G^a_{\mu\nu}$ the QCD field strength
tensor, $Q_{3L}$ the left-handed third generation quark doublet, and $t_R$ and
$b_R$ the right-handed top and bottom singlets, respectively.  Furthermore,
$Y_t$ and $Y_b$ are the top and bottom Yukawa couplings, $f^{abc}$ the $SU(3)$
structure constants and $\tilde\Phi= i\sigma_2\Phi^*$, where $\sigma_2$ is the
second Pauli matrix. Note that $\Lcal_6$ does not represent the complete list of
dimension six operators contributing to Higgs+jet observables. As stated in the
introduction we are mainly interested in the contributions from $O_{3g}$ and
only include the other operators for illustrative purposes.

The main effect of $O_t$ and $O_b$ is to modify the $Ht\bar t$ and $Hb\bar b$
vertex, respectively. The corresponding coupling ratios are
\begin{eqnarray}
  \kappa_t &=& \frac{g^\text{SM}_{Ht\bar t}+\delta g^\text{NP}_{Ht\bar t}}
                  {g^\text{SM}_{Ht\bar t}}
           = 1-\frac{v^2}{2\Lambda^2}C_t 
  \eqsep,\eqsep \nn \\
  \kappa_b &=& \frac{g^\text{SM}_{Hb\bar b}+\delta g^\text{NP}_{Hb\bar b}}
                  {g^\text{SM}_{Hb\bar b}}
           = 1-\frac{v^2}{2\Lambda^2}C_b
  \eqsep,
\end{eqnarray}
where $v=\unit{246}{GeV}$ is the Higgs vacuum expectation value,
$g^\text{SM}_{Ht\bar t}$ and $g^\text{SM}_{Hb\bar b}$ denote the SM values of
the $Ht\bar t$ and $Hb\bar b$ couplings and $\delta g^\text{NP}_{Ht\bar t}$ and
$\delta g^\text{NP}_{Hb\bar b}$ are the shifts in these couplings due to
contributions from the operators $O_t$ and $O_b$, respectively.

The operator $O_g$ generates couplings of one or two Higgs bosons to two, three
or four gluons.  Normalising the $Hgg$ coupling in the infinite top mass limit
to the corresponding SM coupling leads to the coupling ratio
\begin{equation}
  \kappa_g = \frac{g^\text{SM,eff}_{Hgg}+\delta g^\text{NP}_{Hgg}}
                  {g^\text{SM,eff}_{Hgg}}
           = 1 + \frac{12\pi v^2}{\alpha_s\Lambda^2} C_g
  \eqsep.
\end{equation}
Here $g^\text{SM,eff}_{Hgg}$ denotes the effective SM $Hgg$ coupling in the
infinite top mass limit and $\delta g^\text{NP}_{Hgg}$ the coupling induced by
the operator $O_g$.

The operator $O_{3g}$ generates three and four gluon vertices with a modified
momentum dependence and additional vertices with up to six gluons. We do not
introduce a coupling ratio for this operator.

Expected bounds on the Wilson coefficients $C_g$, $C_t$ and $C_b$ can be derived
from the bounds on $\kappa_t$, $\kappa_b$ and $\kappa_g$ given in
\cite{Grojean:2013nya}. Using approximate limits of $1\pm 0.2$ on $\kappa_t$ and
$\kappa_b$ and of $1\pm 0.1$ on $\kappa_g$ as indicated by
their Fig.~2 we find for $\Lambda=\unit{1}{TeV}$ that
\begin{equation}
  |C_t|,|C_b|\lesssim 6.6
  \eqsep,\eqsep
  |C_g|\lesssim 0.0052
  \eqsep.
\end{equation}
Bounds on $C_{3g}$ were estimated in \cite{Cho:1994yu} for a \unit{14}{TeV}
LHC at an integrated luminosity of \unit{30}{\power{fb}{-1}}. Translating
their Fig.~7 into our conventions we find for $\Lambda=\unit{1}{TeV}$
\begin{equation}
  |C_{3g}|\lesssim 0.12
  \eqsep.
\end{equation}
However, it should be noted that their estimate only accounts for statistical
uncertainties. It does not include parton shower effects or other uncertainties,
e.g. from parton distribution functions. The actual bound might therefore be
somewhat weaker.

\begin{figure}
  \centering\includegraphics[scale=0.7]{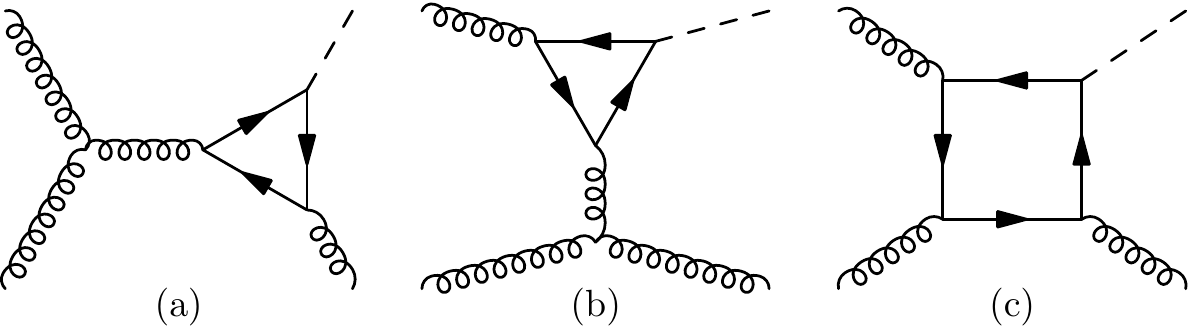}
  \caption{Representative diagrams for $gg\to Hg$ in the SM.}
  \label{fig:SMdiag}
\end{figure}
\begin{figure}
  \centering\includegraphics[scale=0.7]{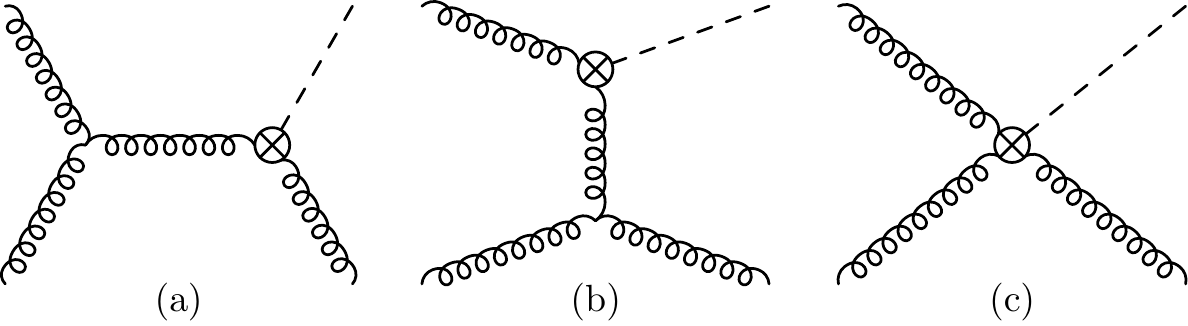}
  \caption{Representative diagrams for $gg\to Hg$ mediated by $O_g$.}
  \label{fig:OGdiag}
\end{figure}

Let us now turn to Higgs production in association with one jet.  In the SM the
$gg\to Hg$ reaction proceeds mainly through top-loop diagrams like the ones
shown in Fig.~\ref{fig:SMdiag}. Interference between top and bottom loop
diagrams leads to corrections of a few percent while the square of the
bottom-loop contributions are negligible. If the effective operators $O_t$ or
$O_b$ are present they simply scale the top-loop contributions and top-bottom
interference by $\kappa_t^2$ and $\kappa_t\kappa_b$, respectively. Due to the
smallness of the top-bottom interference contribution they can not modify the
shape of kinematic distributions in a noticeable way. However, the dimension six
couplings generated by $O_g$ and $O_{3g}$ have a non-trivial momentum dependence
which can lead to very different shapes in kinematic distributions.

\begin{figure*}
  \includegraphics{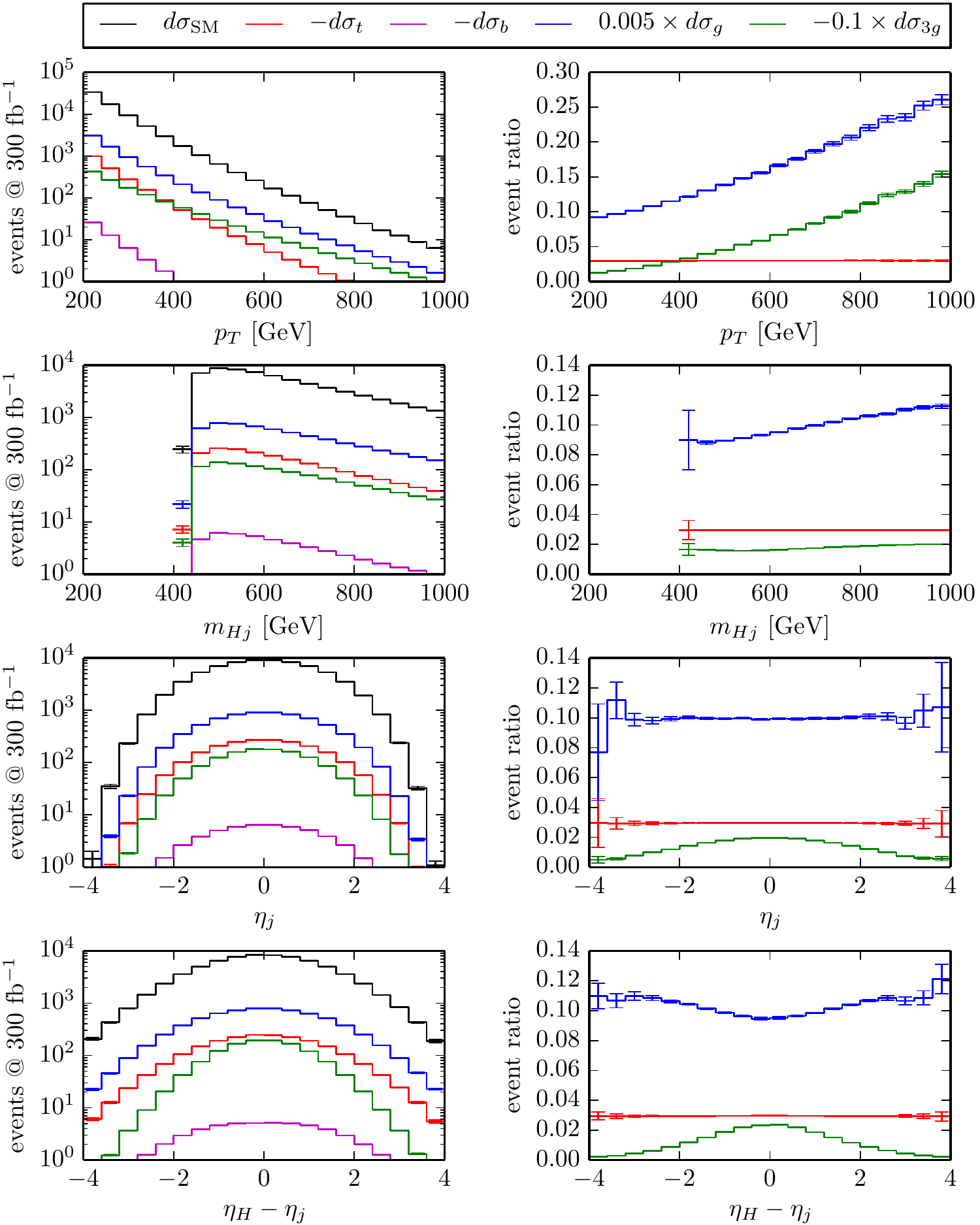}
  \caption{Kinematic distributions due to the differential cross sections
    defined in \eqref{eq:expansion} for $\Lambda=\unit{1}{TeV}$ and a
    \unit{14}{TeV} LHC. A lower $p_T$ cut of \unit{200}{GeV} was applied in all
    plots. Note the sign flips introduced to plot on a logarithmic scale and the
    scaling of $d\sigma_g$ and $d\sigma_{3g}$! The left column shows event
    counts for an integrated luminosity of \unit{300}{\power{fb}{-1}}. The right
    column shows event ratios normalised to the contribution from
    $d\sigma_\text{SM}$. See text for further details.}
  \label{fig:dist}
\end{figure*}

The couplings generated by $O_g$ lead to tree-level diagrams like the ones shown
in Fig.~\ref{fig:OGdiag}. As discussed in \cite{Banfi:2013yoa, Azatov:2013xha,
  Grojean:2013nya}, these contributions modify the shape of the transverse
momentum distribution in $gg\to Hg$ and can be used to discriminate between
$\kappa_t$ and $\kappa_g$ (or equivalently $C_t$ and $C_g$). The operator
$O_{3g}$ modifies the momentum dependence of the three gluon vertex.  It
therefore contributes to $gg\to Hg$ through diagrams like
Fig.~\ref{fig:SMdiag}(a) and (b). The effect of these contributions on kinematic
distributions in $gg\to Hg$ have not yet been discussed in the literature and
are the main result of our analysis.

\section{Kinematic Distributions}
\label{sec:dist}

Let $d\sigma$ denote the gluon fusion contribution to the differential
hadronic cross section for Higgs production in association with one jet in the
presence of the effective operators \eqref{eq:op}. We linearise $d\sigma$
in the Wilson coefficients from \eqref{eq:L6} and write
\begin{equation}\label{eq:expansion}
  d\sigma = d\sigma_\text{SM} + C_t d\sigma_t + C_b d\sigma_b
    + C_g d\sigma_g + C_{3g} d\sigma_{3g}
  \eqsep.
\end{equation}
Note that, since we expand in dimensionless Wilson coefficients,
$d\sigma_t$, $d\sigma_b$, $d\sigma_g$, and $d\sigma_{3g}$ depend on the
cutoff scale $\Lambda$ and scale as $\Lambda^{-2}$.

We have computed the differential cross sections $d\sigma_\text{SM}$,
$d\sigma_t$, $d\sigma_b$, $d\sigma_g$, and $d\sigma_{3g}$ using the full
one-loop result for the top and bottom loops (see Fig.~\ref{fig:SMdiag}). A
number of tools were used for the computation. The Feynman rules for the
effective operators were generated with FeynRules 2.0 \cite{Alloul:2013bka} and
diagrams were generated with FeynArts 3.8 \cite{Hahn:2000kx}. The analytical
computations were done with the Mathematica package HEPMath
\cite{Wiebusch:2014qba}. The results were cross-checked with FormCalc 8.3
\cite{Hahn:1998yk, Hahn:2006zy} and one-loop tensor integrals were computed
numerically with LoopTools 2.9 \cite{Hahn:1998yk}. We used leading order parton
distribution functions (PDFs) from CTEQ (CTEQ6L1) \cite{Pumplin:2002vw} via the
LHAPDF interface. The factorisation and renormalisation scales were set to the
top mass.

The left column of Fig.~\ref{fig:dist} shows histograms of the differential
cross sections $d\sigma_\text{SM}$, $d\sigma_t$, $d\sigma_b$, $d\sigma_g$, and
$d\sigma_{3g}$ for $\Lambda=\unit{1}{TeV}$ as functions of different kinematic
variables. Specifically, we show distributions of the transverse momentum $p_T$
of the jet, the invariant mass $m_{Hj}$ of the Higgs boson and the jet, the
rapidity $\eta_j$ of the jet and
\begin{equation}
  \Delta\eta = \eta_H-\eta_j
  \eqsep,
\end{equation}
where $\eta_H$ is the rapidity of the Higgs boson. A lower $p_T$ cut of
\unit{200}{GeV} was applied in all cases. For convenience the cross sections in
each bin were converted to event counts assuming an integrated luminosity of
\unit{300}{\power{fb}{-1}}. Note, however, that no parton shower or detector
effects are included in our analysis.  To better visualise the different shapes
of the kinematic distributions generated by the operators $O_g$ and $O_{3g}$ we
divided $d\sigma_t$, $d\sigma_g$, and $d\sigma_{3g}$ by $d\sigma_\text{SM}$ and
plot the ratios in the right column of Fig.~\ref{fig:dist}. The error bars only
contain the numerical integration error. No uncertainties due to statistics,
PDFs, luminosity etc.\ are included. Note that $d\sigma_t$, $d\sigma_b$,
$d\sigma_g$ and $d\sigma_{3g}$ are scaled with factors $-1$, $-1$, $0.005$ and
$-0.1$, respectively.  The scales were chosen to be of the same order as the
expected bounds on the Wilson coefficients, as discussed in
Sec.~\ref{sec:op}. The signs were chosen in order to be able to plot on a
logarithmic scale.

From the right column of Fig.~\ref{fig:dist} we see that $O_t$ does not modify
the shape of kinematic distributions in a noticeable way. Theoretically, a small
variation should be present due to the top-bottom interference contribution to
$d\sigma_\text{SM}$, but this is too small to be visible in the figure. The
relative size of the contribution from $O_g$ grows larger with large $p_T$, in
accordance with the findings of \cite{Banfi:2013yoa, Azatov:2013xha,
  Grojean:2013nya}. In addition we see that $m_{Hj}$ and $\Delta\eta$ are also
useful for discriminating between $O_g$ and $O_t$. Note that only the
\emph{shape} of the ratios shown in the right column of Fig.~\ref{fig:dist} is
relevant for distinguishing the different operators. In particular, a constant
offset of the ratio curve for, say, $O_g$ can not be distinguished from a
contribution from $O_t$.

From the top-right plot of Fig.~\ref{fig:dist} we see that the $p_T$
distribution due to $O_{3g}$ is a linear combination of the distributions due to
$O_g$ and $O_t$. Using only $p_T$ distibutions (and treating $C_t$ as a free
parameter in the fit) it is therefore not possible to distinguish between $O_g$
and $O_{3g}$. This new degeneracy can be broken by considering other kinematic
variables such as $\eta_j$ and $\Delta\eta$.  In particular, the ratio curve for
$O_g$ in the variable $\eta_j$ is completely flat while the one for $O_{3g}$ is
peaked around $\eta_j=0$. Disregarding constant offsets the variations of the
ratio curves for $O_{3g}$ are of the same size as the variations for $O_g$ if
the Wilson coefficients of both operators are of the order of the expected
bounds. Thus the possible impact of $O_{3g}$ on Higgs+jet observables is as
large as that of $O_g$.

\section{Conclusions}
\label{sec:dist}

Higgs production in association with a jet receives contributions not only from
effective dimension six operators involving the Higgs doublet but also from the
triple gluon operator $O_{3g}$. In this paper we have calculated the
differential cross section for $gg\to Hg$ in the presence of the dimension six
operators $O_{3g}$, $O_g$, $O_t$ and $O_b$. To the best of our knowledge the
contributions due to $O_{3g}$ have not yet been discussed in the literature. We
find that the distributions of the transverse momentum $p_T$ of the jet are
insufficient to discriminate between $O_g$ and $O_{3g}$ if the Wilson
coefficient of $O_t$ is free to float in the fit. This new degeneracy can be
broken by considering distributions in the jet rapidity $\eta_j$ and the
difference $\Delta\eta$ between the Higgs and jet rapidity. In particular, the
shape of the $\eta_j$ distribution is only modified by $O_{3g}$ but not by $O_g$
or $O_t$. The distortions due to $O_{3g}$ are of the same size as the
distortions due to $O_g$ if the Wilson coefficients of both operators are of the
order of the expected bounds. Consequently, a full model independent analysis of
new physics in Higgs+jet observables must take the contributions from $O_{3g}$
into account.

\subsubsection*{Acknowledgements}
This work is supported by the European Research Council under the European
Union's Seventh Framework Programme (FP/2007-2013) / ERC Grant Agreement
n.279972. DG thanks Michael Spannowsky for hospitality in the IPPP, Durham
University where this project was envisaged, and Roberto Contino for discussions.

\input{higgs+jet.bbl}
\end{document}

%% file: higgs+jet.bbl
%